\documentclass[sn-aps]{sn-jnl}

\usepackage{graphicx}%
\usepackage{multirow}%
\usepackage{amsmath,amssymb,amsfonts}%
\usepackage{amsthm}%
\usepackage{mathrsfs}%
\usepackage[title]{appendix}%
\usepackage{xcolor}%
\usepackage{textcomp}%
\usepackage{manyfoot}%
\usepackage{booktabs}%
\usepackage{algorithm}%
\usepackage{algorithmicx}%
\usepackage{algpseudocode}%
\usepackage{listings}%
\usepackage{textcomp}%
\usepackage{subfloat}%
\usepackage{caption}%
\usepackage{subcaption}%

\theoremstyle{thmstyleone}%
\theoremstyle{thmstyletwo}%

\theoremstyle{thmstylethree}%

\begin{document}

\title[Article Title]{Preparation of Epitaxial Scandium Trifluoride Thin Films using Pulsed Laser Deposition }


\author*[1,2]{\fnm{Amani S.} \sur{Jayakody}}\email{amani.jayakody@uconn.edu}

\author[1]{\fnm{Joseph} \sur{Budnick}}\email{budnick@phys.uconn.edu}

\author[1,3]{\fnm{Jason N. } \sur{Hancock}}\email{jason.hancock@uconn.edu}

\author[3]{\fnm{Daniela} \sur{Morales}}\email{daniela.morales@uconn.edu}

\author[1,3]{\fnm{Barrett} \sur{ O. Wells}}\email{barrett.wells@uconn.edu}

\affil[1]{\orgdiv{Department of Physics}, \orgname{University of Connecticut} \orgaddress{\street{}, \city{Storrs}, \postcode{06269}, \state{CT}, \country{USA}}}

\affil[2]{\orgdiv{Department of Physics}, \orgname{Hamilton College} \orgaddress{\street{}, \city{Clinton}, \postcode{13323}, \state{NY}, \country{USA}}}

\affil[3]{\orgdiv{Institute of Materials Science}, \orgname{University of Connecticut}, \orgaddress{\street{}, \city{Storrs}, \postcode{06269}, \state{CT}, \country{USA}}}


\abstract{Bulk Scandium trifluoride ($\mathrm{ScF_3}$) is known for a pronounced negative thermal expansion (NTE) over a wide range of temperature, from $10 \mathrm{K}~\text{to}~ 1100~\mathrm{K}$. The structure of $\mathrm{ScF_3}$ can be described as an $\mathrm{ABX_3}$ perovskite with an empty A-site and a space group of Pm-3m. Growing thin films of $\mathrm{ScF_3}$ allows for tuning the lattice constant, the thermal expansion, and the construction of devices based upon differential thermal expansion. We have investigated the growth of $\mathrm{ScF_3}$ films on oxide and fluoride substrates using pulsed laser deposition (PLD) This letter describes the successful growth recipe for producing high quality epitaxial $\mathrm{ScF_3}$ thin films on positive thermal expansion (PTE) lithium fluoride ($\mathrm{LiF}$) substrates, at substrate temperature, $350^{\circ}\mathrm{C}$ with a laser repetition rate of $1~\mathrm{Hz}$, with an energy per pulse of $600~\mathrm{mJ}$, under a vacuum of $1.5\times 10^{-6}~ \mathrm{torr}$, for a growth time of $6$ hours. However, even for films with excellent epitaxy and sharp peaks along the principal axes, diffraction peaks from certain crystallographic directions are extremely broad, with the example of ($104$) reflections, in this work. We attribute this broadening to disorder in the $\mathrm{F_6}$ octahedral rotations that occur as an attempt to accommodate the large temperature-induced lattice mismatch that results in cooling from the growth temperature for this system of a NTE film mated to a PTE substrate. }

\maketitle

\section{Introduction}\label{sec1}

Usually, materials in nature expand when they get heated, but there are some materials which shrink upon heating; this phenomenon is called negative thermal expansion (NTE). NTE is common in polymers and biomolecules, and as well as in inorganic systems such as Zirconium Tungstate ($\mathrm{ZrW_2O_8}$) and scandium trifluoride ($\mathrm{ScF_3}$) which are now proven to be stemming from the entropic elasticity of an ideal, freely jointed chain \cite{b1}. This exceptional phenomenon attracts attention as it has potential for applications. Positive thermal expansion (PTE) of materials causes many problems in engineering and as well as in day-to-day life. NTE materials are most of the time put together with PTE materials in order to make composite materials with zero thermal expansion.  Zero thermal expansion materials are very much useful in electronics \cite{b2}. $\mathrm{ScF_3}$ is a cubic nonmagnetic, ionic insulator and one of the very few materials which exhibit NTE over a wide temperature range, from $10-1100~\mathrm{K}$ \cite{b3}. In addition, for much of the range the effect is large with a volume thermal expansion coefficient below room temperature of $\mathrm{\alpha}_{V} =-34 \times 10^{-6}~\mathrm{K^{-1}}$. It has a room temperature lattice constant of $4.0127 ~\mathrm{\AA}$. The structure of $\mathrm{ScF_3}$ is that of the cubic perovskite ($\mathrm{ABX_3}$) with an empty A site. $\mathrm{ScF_3}$ remains cubic over the entire temperature range, with a barely avoided transition to a rhombohedral phase near $0~\mathrm{K}$ \cite{b4}. 
Our interest is to see what happens to film-bonded materials with positive and negative thermal expansion. While there may be technological interests (bimetallic strips), there is a fundamental question about how a film can accommodate stress/strain, with perhaps notable differences if that stress builds with temperature versus at the initial growth or introduced suddenly at a phase transition. In $\mathrm{ScF_3}$ thin films we expect a very large, induced strain to develop as a function of temperature, regardless of the thickness of the film. This is a good chance to study the effects of such large thermally induced strains and extreme issues of strain engineering. In addition, $\mathrm{ScF_3}$ has been described as hosting a structural quantum phase transition (or just barely avoiding such a transition) and the large strains may induce the interesting effects and a new transition \cite{b4}. In this letter, we report the successful growth routes of synthesizing high quality, epitaxial $\mathrm{ScF_3}$ thin films, with an explanation to the unusual broadening in partially out-of-plane film peaks suggesting some disorder in the $\mathrm{ScF_3}$ structure.
\section{Experimental}\label{sec2}
\begin{figure}[h]
\centering
\includegraphics[scale=0.4]{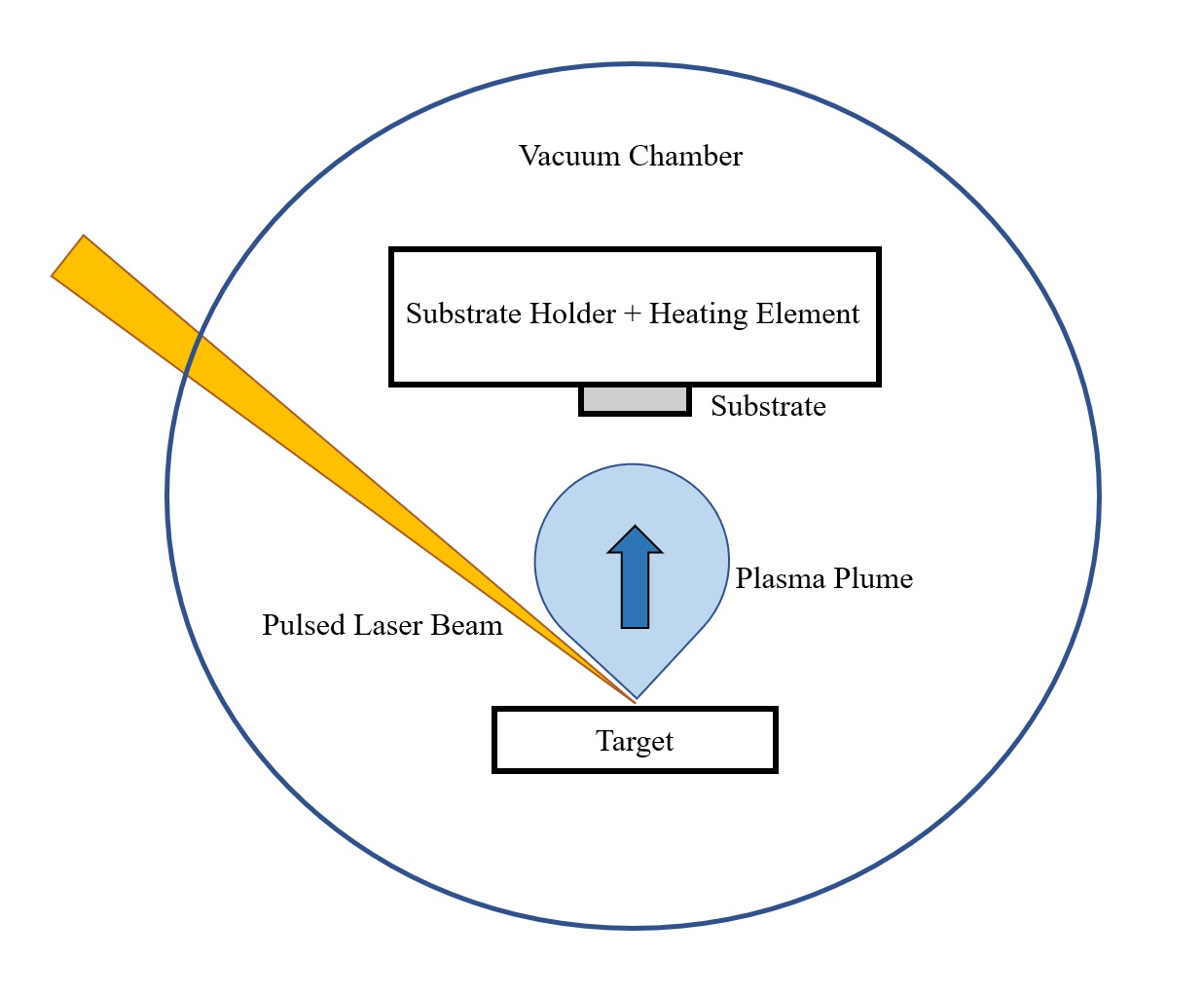}
\caption{Schematic illustration of a PLD system for the deposition of thin films.}
\label{f1}
\end{figure}

We attempted to grow quality epitaxial $\mathrm{ScF_3}$ thin films on different oxide and fluoride substrates by pulsed laser deposition (PLD). PLD is typically done with UV photons to promote energy absorption without heating/melting, and we use a $\mathrm{KrF}$ excimer laser beam with $248~\mathrm{nm}$ wavelength, to be focused inside a vacuum chamber to strike a target of the material that is to be deposited (Fig.~\ref{f1}).
The initial powder and out-of-plane x-ray diffraction (XRD) measurements were performed using a two-circle powder diffractometer (Bruker D2phaser). It is a very fast and convenient measurement that can be performed initially, to confirm the structure of the target and the thin films grown. The epitaxy of the thin films was further checked using a four-circle diffractometer with a circular area detector  (Oxford XCalibur \textsuperscript{TM} PX Ultra). The 4-circle kappa goniometer allows easy crystal mounting and alignment. The area detector makes it easy to investigate the epitaxy and texture of the material. A Rigaku SmartLab x-ray diffractometer was used for further structural analysis of the thin films, as it allows for phi-rotation scans at grazing incidence (in-plane) to test the epitaxy and in-plane texture.  This diffractometer has the ability to tilt the 2theta arm out of the vertical scattering plane to allow for easy alignment for grazing incident experiments. All x-ray measurements were made with $\mathrm{Cu~K_{\alpha}}$ radiation of wavelength $1.5406~\mathrm{\AA}$.
\section{Results and Discussion}\label{sec3}
\begin{figure}[h]
	\centering
	\includegraphics[scale=0.65]{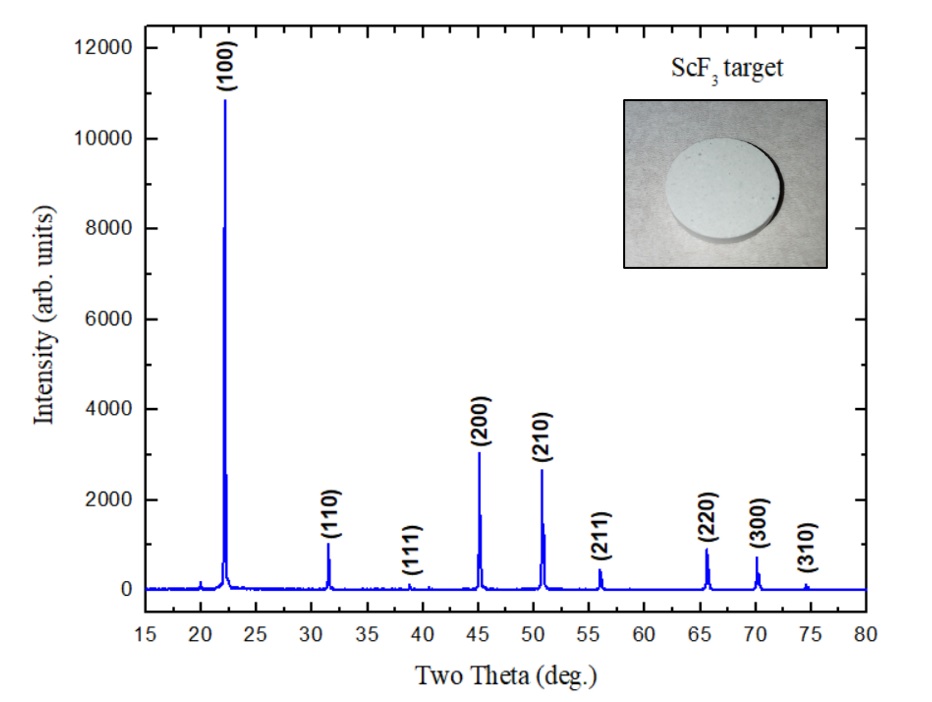}
	\caption{X-ray diffractogram and the appearance (inset) of the $\mathrm{ScF_3}$ target. Note that the peaks that are not listed are coming from the $\mathrm{Cu~K_{\beta}}$ radiation produced by the x-ray source.}
	\label{f2}
\end{figure}

A commercially made $19~\mathrm{mm}$ diameter and $4~\mathrm{mm}$ thickness, vacuum pressed $\mathrm{ScF_3}$ target (Demaco Vacuum) with $99.5\%$ purity was used in the PLD process. The stability and the structure of the $\mathrm{ScF_3}$ target was confirmed by performing powder XRD measurements using the powder diffractometer where the two-axis scattering of q-space is normal to the sample surface. [Fig. \ref{f2}]. $\mathrm{ScF_3}$ target is white in appearance as shown in Fig.\ref{f2} inset. Initially, we attempted to grow $\mathrm{ScF_3}$ thin films on oxide substrates such as strontium titanate ($\mathrm{SrTiO_3}$) and magnesium oxide ($\mathrm{MgO}$). $\mathrm{SrTiO_3}$ has a perovskite cubic structure with the space group Pm-3m and has a lattice constant of $3.905~\mathrm{\AA}$. $\mathrm{MgO}$ has a cubic structure with the space group Fm-3m, and has a lattice constant of $4.2112~\mathrm{\AA}$. The films were grown at different substrate temperatures ranging from $25-700~^{\circ}\mathrm{C}$, with a laser voltage of $27~\mathrm{kV}$ and a repetition rate of $1-2~\mathrm{Hz}$ under a chamber vacuum of $1.5\times10^{-6}~\mathrm{torr}$, for a growth time of 1-6 hours.  However, for any growth conditions, the diffraction pattern, normal to the film surfaces, taken using the powder diffractometer did not show any $\mathrm{ScF_3}$ peaks on both $\mathrm{SrTiO_3}$ and $\mathrm{MgO}$ substrates. It doesn't necessarily mean that there is no film at all, but it implies that there isn't sufficient crystalline/epitaxial film to get any X-ray diffraction. This may be due to the relatively high lattice mismatch and the chemical incompatibility between the fluoride thin film and the oxide substrates. The lattice mismatch between $\mathrm{ScF_3}$ and $\mathrm{SrTiO_3}$ is $2.68\%$ and the lattice mismatch between $\mathrm{ScF_3}$ and $\mathrm{MgO}$ is  $-4.95\%$. 

\begin{figure}
	\centering
	\includegraphics[scale=0.41]{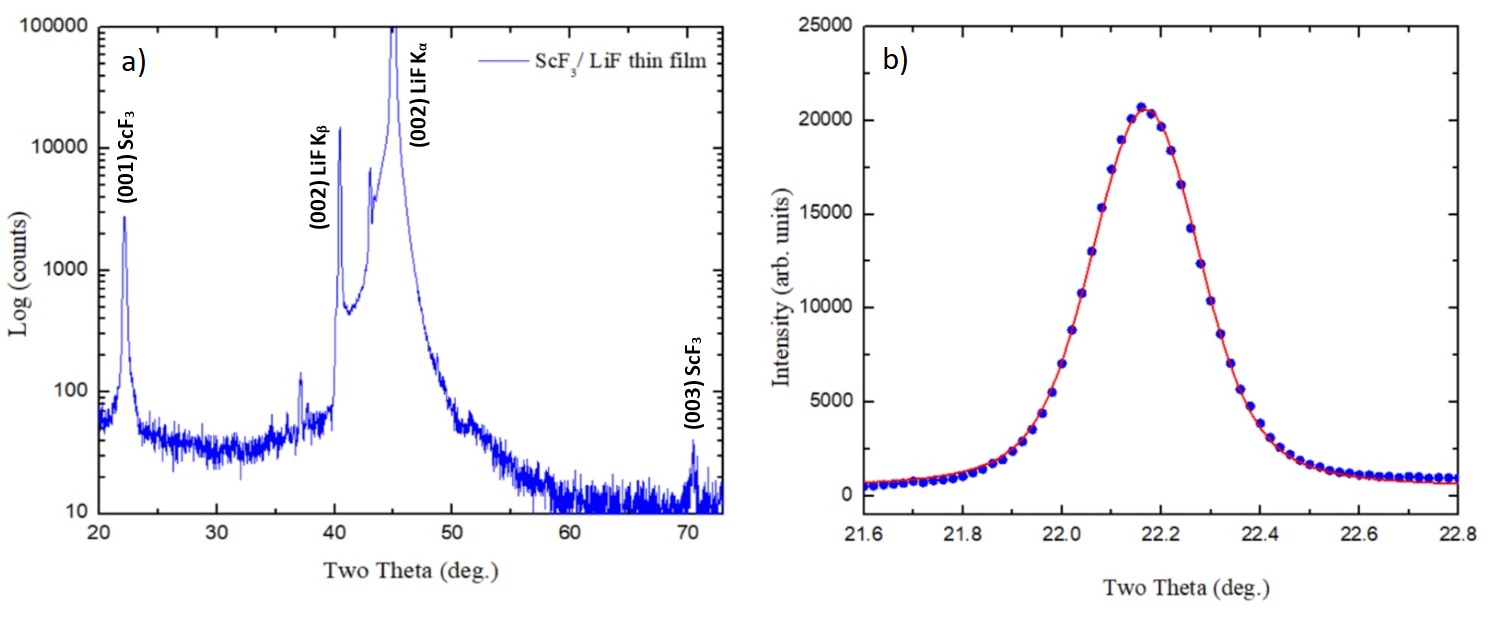}
	\caption{a) Full Out-of-plane XRD diffractogram for $\mathrm{ScF_3/LiF}$ thin film of thickness $30~\mathrm{nm}$. b) Zoomed in out-of-plane $\mathrm{ScF_3}$ $(001)$ XRD peak with FWHM of $0.20^{\circ}$, where the inherent resolution of the diffractometer being $0.16^{\circ}$. Note that the peaks that are not listed are coming from the $\mathrm{Cu~K_{\beta}}$ radiation produced by the x-ray source, and from features unique to the commercially bought $\mathrm{LiF}$ substrates. }
	\label{f3a}
\end{figure}

As the growth of $\mathrm{ScF_3}$ thin films on oxide substrates was unsuccessful, we decided to move on to fluoride substrates with the hope that better chemical compatibility would enable epitaxial growth. We chose $\mathrm{LiF}$ as a substrate material because $\mathrm{LiF}$ has a cubic structure of Fm-3m space group with a lattice constant of $4.027~\mathrm{\AA}$. The lattice mismatch between $\mathrm{ScF_3}$ and $\mathrm{LiF}$ is $0.355\%$ at room temperature, is an excellent lattice match. As expected, we were able to grow high quality epitaxial $\mathrm{ScF_3}$ thin films on $\mathrm{LiF}$ substrates, due to the excellent lattice match and the chemical compatibility. $\mathrm{LiF}$ has a positive thermal expansion and has a thermal expansion coefficient ($\alpha_V$) of $37\times10^{-6}~\mathrm{K^{-1}}$ below room temperature \cite{b5}. Hence, $\mathrm{LiF}$ is a good candidate to address the question as to how an NTE thin film accommodates stress/strain on a PTE substrate, while maintaining a technological interest in aspects such as bimetallic strips. The optimized growth recipe for high quality, epitaxial $\mathrm{ScF_3}$ thin films on $\mathrm{LiF}$ substrates was found at substrate temperature, $350~^{\circ}\mathrm{C}$ with a laser repetition rate of $1~\mathrm{ Hz}$, with an energy per pulse of $600~\mathrm{mJ}$ (laser voltage of $27~\mathrm{kV}$), under a vacuum of $1.5\times10^{-6}~\mathrm{torr}$, for a growth time of 6 hours. Note that we have used higher growth times in this process, to grow thin films with thicknesses ranging from $20-30~\mathrm{nm}$. The vacuum pressed $\mathrm{ScF_3}$ target was relatively hard to ablate, and the plume produced by the target even with the highest laser energy was very weak. This is presumably because the photon coupling is relatively poor here. $\mathrm{KrF}$ is at $248~\mathrm{nm}$ ($5~\mathrm{eV}$) with a calculated band gap for $\mathrm{ScF_3}$ of greater than $8~\mathrm{eV}$ \cite{b6}. Hence, higher growth times were used when growing the thin films, to get a reasonable thickness for the XRD characterization purpose.
We performed an x-ray scattering analysis of our films on a lab based, Rigaku diffractometer, at room temperature. The lower symmetry of $\mathrm{ScF_3}$ versus $\mathrm{LiF}$ means that approximately half of the film Bragg peaks occur far from any interfering of substrate diffraction, for example $\mathrm{ScF_3}$ $(001)$ peaks are easy to measure whereas $(002)$ peaks can be lost in the substrate scattering (Fig.~\ref{f3a}a). Our best films have sharp out of plane Bragg peaks, with the $(001)$ shown in Fig.~\ref{f3a}b. 

\begin{figure}
	\centering
	\includegraphics[scale=0.46]{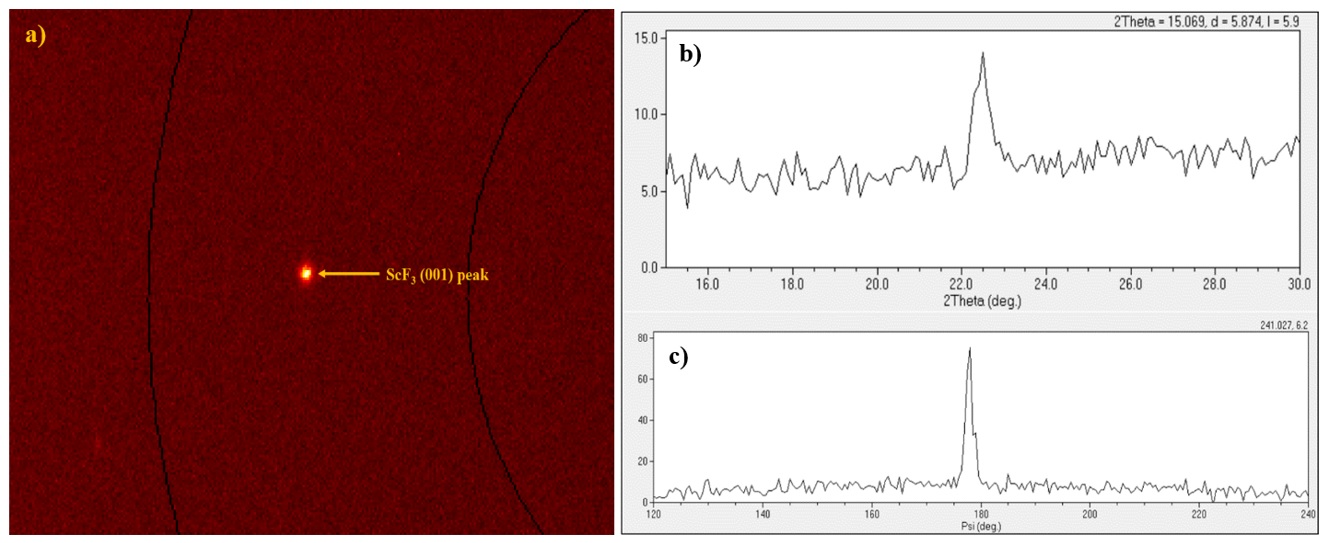}
	\caption{a) XRD image which shows the epitaxy of $\mathrm{ScF_3}$ thin film on $\mathrm{LiF}$ substrate, during a static scan where $\omega=-67^{\circ},~\theta=-30^{\circ},~K=-134^{\circ},~\phi=0^{\circ}$ at a detector distance $65~\mathrm{mm}$. b) Line cuts through the same data set in part a in the two theta (horizontal) direction. The strong peak near $22$ degrees is the $\mathrm{ScF_3}$ $(001)$ film peak. The sample angle $\omega$ is optimized for the film peak. c) Cut in the chi (vertical) direction of the same data set. The large tails represent the powder ring. The peak is about a degree wide. Here, psi is equivalent to chi in the diffractometer.}
	\label{F5}
\end{figure}

Fig.~\ref{F5}a shows a two-dimensional image of the x-ray diffraction pattern for the film grown on $\mathrm{LiF}$ checking the epitaxy of the $\mathrm{ScF_3}$ film, using the four-circle diffractometer with an area detector. Lines of constant two-theta which form parabolas are shown in Fig.~\ref{F5}a as a reference. Two theta increases from right to left. Dispersed along the parabolas, approximately vertically, is the standard scattering angle chi - a rocking curve out of the scattering plane. The image is taken with the in-plane sample angle $\omega$ set to optimize the $\mathrm{ScF_3}$ film $(001)$ peak at the expense of any other $\mathrm{LiF}$ substrate peaks. However, since there are no substrate peaks near the film $(001)$ peak, a clean XRD image of the film peak was obtained. We did not see any powder rings in the XRD image obtained suggesting any partial epitaxy in the film. Fig.~\ref{F5}b and \ref{F5}c show the line cuts through the same data set in Fig.~\ref{F5}a, in the $2\theta$ and chi direction.

\begin{figure}[h]
		\centering
		\includegraphics[scale=0.38]{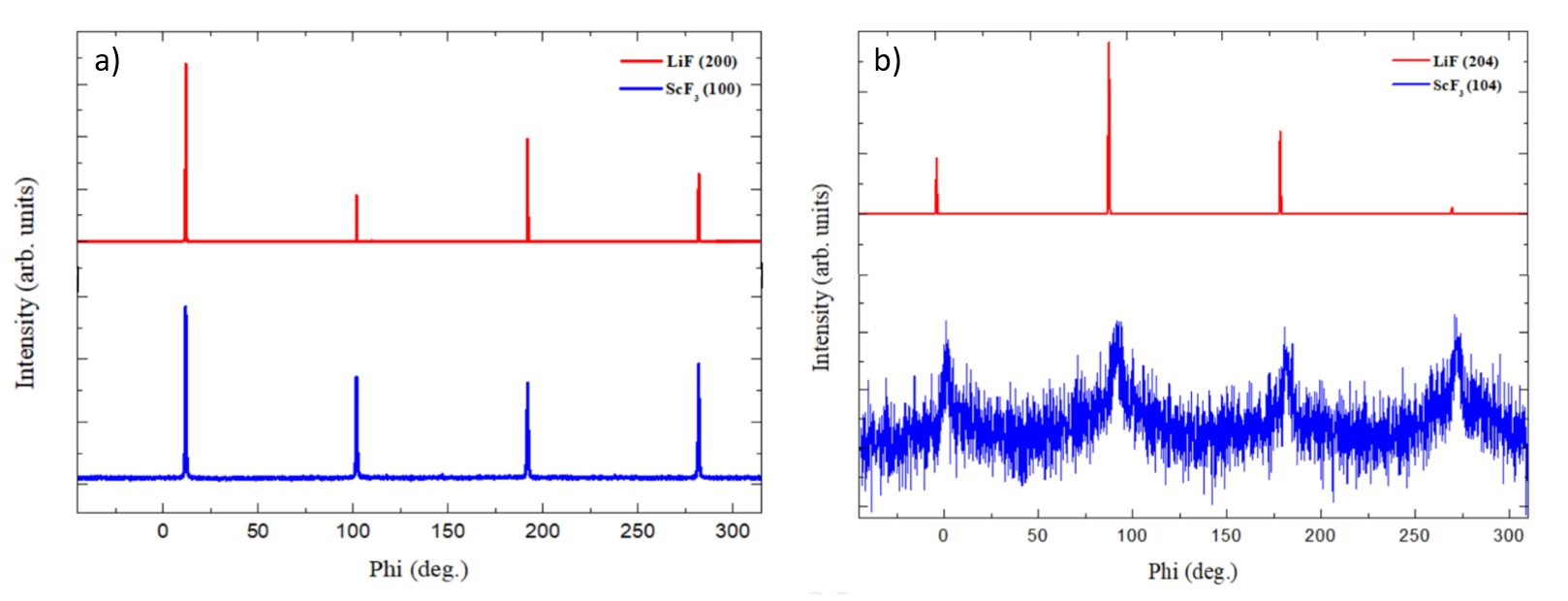}
		\caption{a) Phi scans (rotation around $(001)$ direction) showing 4-fold symmetry for completely in-plane $\mathrm{LiF}$ $(200)$ (FWHM $= 0.22^{\circ}$) and $\mathrm{ScF_3}$ $(100)$ (FWHM $= 0.57^{\circ}$) peaks. b) Phi scans (rotation around $(001)$ direction) showing 4-fold symmetry for mostly out-of-plane $\mathrm{LiF}$ $(204)$ (FWHM $= 0.62^{\circ}$) and $\mathrm{ScF_3}$ $(104)$ (FWHM $= 10.86^{\circ}$). Note that the y-axis is linear in both a) and b).}
		\label{f5a}
	\end{figure}

Next, to further confirm the epitaxy of the films, Rigaku diffractometer was used for grazing incident measurements. In-plane $(100)$ peaks are also sharp, measured in grazing incidence, with rotation about the substrate normal showing excellent epitaxy with the films axes perfectly aligned with the substrate (Fig.~\ref{f5a}). However, we have also found a surprising result in measuring the width and alignment of the $(104)$ Bragg peak with respect to the substrate $(204)$ peak, as shown in Fig.~\ref{f5a}b. This peak is also well aligned with the substrate, but extremely broad with respect to rotation around the substrate normal. The full width at half maximum (FWHM) of the partially out-of-plane $(104)$ film peak is $10.86^{\circ}$, which is about 18 times the FWHM of $(204)$ substrate peak, $0.62^{\circ}$. In addition, there are notably extended peak tails as is obvious from the $\mathrm{ScF_3}$ $(104)$ trace in Fig.~\ref{f5a}b. Under typical circumstances, such broad peaks would be due to poor epitaxy. In films this is described as a broad in-plane film texture and in crystals as mosaic spread around the $[001]$ axis. However, that cannot be the case here as the primary $(100)$ peaks are sharp and well aligned.

As mentioned in the previous paragraph, the phi scans in Fig.~\ref{f5a}a and~\ref{f5a}b are rotations of the sample around the $[001]$ direction. The $(100)$ peak (Fig.~\ref{f5a}) is a true transverse scan and indicates that the mosaic spread around the $[001]$ is $0.576^{\circ}$ or less. Similarly, Fig.~\ref{F5}a,~\ref{F5}b, and~\ref{F5}c indicates that the mosaic spread around an in-plane axis is also small ($4.84^{\circ}$). Thus, the $10.86^{\circ}$ broad phi scan for the $(104)$ peak and extended peak-tails cannot be attributed to mosaic or film texture in the traditional sense, but rather must indicate disorder induced in $(104)$ Bragg plane alignment that is separate from the principal cubic planes. These apparently incompatible phenomena coexist because of the unique phonon properties of $\mathrm{ScF_3}$ and the dual structural constraints in epitaxial films.

Perovskites are well known to feature a variety of cation cage rotations. In the case of $\mathrm{ScF_3}$, $\mathrm{F_6}$ octahedral rotations about $(100)$ directions are associated with the negative thermal expansion effect, though other rotation axes may also lead to NTE \cite{b7}.  Inelastic x-ray scattering of bulk single crystal $\mathrm{ScF_3}$ at ambient pressure showed considerable softening along the entire branch from the M to R points along the Brillouin zone edge  \cite{b4}. These zone-edge phonon branches represent dynamical staggered rotations of the octahedral $\mathrm{F_6}$ cage. Further, in related compounds and in $\mathrm{ScF_3}$ under hydrostatic pressure, the cubic structure collapses to a lower symmetry rhombohedral unit cell associated with $\mathrm{F_6}$ cage rotations about the $[111]$ direction - the R point phonon mode \cite{b4}, \cite{b8}. This leads to a picture of unique behavior in our $\mathrm{ScF_3}$ film driven by the combination of the clamping effect due to the film-substrate bonding forces and the biaxial strain induced in the $\mathrm{ScF_3}$ film by the positive thermal expansion $\mathrm{LiF}$ substrate as it cools from the growth temperature. The former effect forces a uniform distribution and alignment of the primary cubic crystalline planes and suppresses the tendency to form the lower symmetry rhombohedral phase. However, to accommodate the compression, $\mathrm{F_6}$ octahedra must rotate. The equally soft modes along M to R lead to a variety of differing rotations and thus the considerable disorder measured. 

\section{Conclusion}\label{sec4}
We have fabricated $\mathrm{ScF_3}$ thin films on various oxide and fluoride substrates using PLD, and we were able to synthesize high quality epitaxial NTE $\mathrm{ScF_3}$ thin films on PTE $\mathrm{LiF}$ substrates. The optimized growth conditions were found at substrate temperature, $350^{\circ}~\mathrm{C}$, a laser repetition rate of $1~\mathrm{Hz}$, an energy per pulse of $600~\mathrm{mJ}$ (laser voltage of $27~\mathrm{kV}$), under a vacuum of $1.5\times10^{-6}~\mathrm{torr}$, for a growth time of 6 hours.  $\mathrm{LiF}$ substrates were used as the substrate material due to the excellent chemical compatibility and the lattice match. Even our best films with good uniformity and epitaxy as measured by the principle diffraction peaks, exhibit unusual broadening of the $(104)$ diffraction peaks for rotations around the $[001])$ axis perpendicular to the film-substrate interface. We postulate that this surprising behavior is associated with how the film adjusts to accommodate the increasing compressive strain that occurs while the sample is cooled from growth to room temperature, with the film's tendency to expand is frustrated by the contraction of the much larger substrate. The frozen disorder would be a natural consequence of the strain freezing out different rotation modes from the line of soft modes identified in previous resonant x-ray studies. We expect this broadening will increase as the temperature is lowered further, which opens an avenue for further studies.






\end{document}